\documentclass[prd,preprintnumbers,superscriptaddress,nofootinbib,eqsecnum]{revtex4}

\usepackage{color}
\usepackage{calc}
\usepackage{amsmath,amssymb,graphicx}
\usepackage{tensor}
\usepackage{bm}
\usepackage{svn-multi}
\usepackage[colorlinks, pdfborder={0 0 0}]{hyperref}
\usepackage{url}
\definecolor{LinkColor}   {rgb}{0.75, 0, 0}
\definecolor{CiteColor}   {rgb}{0, 0.75, 0}
\definecolor{UrlColor}    {rgb}{0, 0, 0.75}
\hypersetup{linkcolor=LinkColor}
\hypersetup{citecolor=CiteColor}
\hypersetup{urlcolor=UrlColor}

\numberwithin{equation}{section}
\newcommand{\Ys}{{{}^{-s}Y}}
\newcommand{\Ytwo}{{{}^{-2}Y}}
\newcommand{\tens}[1]{\mytensor{#1}}
\newcommand{\xhat}{\vec{e}_x}
\newcommand{\yhat}{\vec{e}_y}
\newcommand{\zhat}{\vec{e}_z}
\newcommand{\ihat}{\vec{e}_i}
\newcommand{\jhat}{\vec{e}_j}
\newcommand{\rhat}{\vec{e}_{r}}
\newcommand{\iotahat}{\vec{e}_{\iota}}
\newcommand{\phihat}{\vec{e}_{\phi}}
\newcommand{\eplus}{\tens{e}_+}
\newcommand{\ecross}{\tens{e}_\times}

\newcommand{\Flux}{\mathcal{F}}
\newcommand{\coloneq}{\mathrel{\mathop:}=}
\newcommand{\define}{\coloneq}

\newcommand{\barie}{{\scriptscriptstyle \hspace{-0.55em} \mathbf{\stackrel{-}{}}}}

\svnidlong 
{$HeadURL: https://www.ninja-project.org/svn/ninja2/DataFormatDoc/NRDataFormat.tex $} 
{$LastChangedDate: 2011-02-24 21:39:12 +0000 (Thu, 24 Feb 2011) $} 
{$LastChangedRevision: 585 $} 
{$LastChangedBy: sascha $} 
\svnid{$Id: NRDataFormat.tex 585 2011-02-24 21:39:12Z sascha $}

\begin{document}

\newcommand{\CIT}{Theoretical Astrophysics, California Institute of
  Technology, Pasadena, California 91125, USA}\affiliation{\CIT}
  
\newcommand{\LIGO}{LIGO Laboratory, California Institute of
  Technology, Pasadena, California 91125, USA}\affiliation{\LIGO}

\newcommand{\Cornell}{Center for Radiophysics and Space
    Research, Cornell University, Ithaca, New York, 14853}\affiliation{\Cornell}

\newcommand{\Syracuse}{Department of Physics, Syracuse University, Syracuse, NY 13244}\affiliation{\Syracuse}

\newcommand{\CU}{Cardiff University, Cardiff, CF2 3YB, United
  Kingdom}\affiliation{\CU} 

\newcommand{\AEI}{Max-Planck-Institut
  f\"ur Gravitationsphysik, Albert-Einstein-Institut, Am M\"uhlenberg
  1, D-14476 Golm, Germany}\affiliation{\AEI}

\newcommand{\CCT}{Center for Computation and Technology, Louisiana
  State University, Baton Rouge, LA 70803, USA}%

\newcommand{\LSU}{Department of Physics and Astronomy, Louisiana
  State University, Baton Rouge, LA 70803, USA}%

\newcommand{\PSU}{The Pennsylvania State University, University Park, PA 
16802, USA}

\newcommand{\UIB}{Departament de F\'isica, Universitat de les Illes
  Balears, Crta. Valldemossa km 7.5, E-07122 Palma, Spain}%

\newcommand{\uwm}{Department of Physics, University of Wisconsin --
  Milwaukee, PO Box 413, Milwaukee, WI 53201, USA}%

\newcommand{\RIT}{Center for Computational Relativity and Gravitation,
  School of Mathematical Sciences, Rochester Institute of Technology,
  Rochester, New York 14623, USA}%

\newcommand{\UM}{Department of Physics and Astronomy, The University
  of Mississippi, University, MS 38677-1848, USA}%

\title{Data formats for numerical relativity}

\author{P.~Ajith}\email{ajith@caltech.edu}\affiliation{\CIT}\affiliation{\LIGO}
\author{M.~Boyle}\email{mob22@cornell.edu}\affiliation{\Cornell}
\author{D.~A.~Brown}\email{dabrown@physics.syr.edu}\affiliation{\Syracuse}
\author{S.~Fairhurst}\email{stephen.fairhurst@astro.cf.ac.uk}\affiliation{\CU}
\author{M.~Hannam}\email{mark.hannam@astro.cf.ac.uk}\affiliation{\CU}
\author{I.~Hinder}\email{ian.hinder@aei.mpg.de}\affiliation{\AEI}
\author{S.~Husa}\email{sascha.husa@uib.es}\affiliation{\UIB}
\author{B.~Krishnan}\email{badri.krishnan@aei.mpg.de}\affiliation{\AEI}
\author{R.~A.~Mercer}\email{ram@gravity.phys.uwm.edu}\affiliation{\uwm}
\author{F.~Ohme}\email{frank.ohme@aei.mpg.de}\affiliation{\AEI}
\author{C.~D.~Ott}\email{cott@tapir.caltech.edu}\affiliation{\CIT}\affiliation{\CCT}
\author{J.~S.~Read}\email{jsread@relativity.phy.olemiss.edu}\affiliation{\UM}
\author{L.~Santamar\'ia}\email{luciasan@caltech.edu}\affiliation{\CIT}\affiliation{\LIGO}
\author{J.~T.~Whelan}\email{john.whelan@astro.rit.edu}\affiliation{\RIT}

\begin{abstract}
  This document proposes data formats to exchange numerical relativity results,
  in particular gravitational waveforms. The primary goal is to
  further the interaction
  between gravitational-wave source modeling groups and the
  gravitational-wave data-analysis community. We present a simple
  and extendible format which is applicable to various kinds of gravitational
  wave sources
  including binaries of compact objects and systems undergoing
  gravitational collapse, but is nevertheless sufficiently general to be
  useful for other purposes.\\
\end{abstract}

\preprint{LIGO-T070072-v3}

\maketitle

\section{Introduction}
\label{sec:intro}

Numerical relativity (NR) has made enormous progress within the last
few years. Following the initial breakthroughs of 2005
\cite{Pretorius:2005gq,Campanelli:2005dd, Baker:2005vv}, a number of
numerical codes are available to perform sufficiently accurate
simulations of the inspiral, merger, and ringdown phases of generic
black-hole-binary systems (for overviews see
e.g.~\cite{Pretorius:2007nq, Husa:2007zz,
  Hannam:2009rd,Hinder:2010vn,Centrella:2010mx}).
Similarly, significant progress has
been made in the numerical simulation of the inspiral, coalescence,
and post-merger dynamics of binaries involving neutron stars, and of
stellar gravitational collapse (see
e.g.~\cite{Faber:2009zz,Duez:2009yz,Ott:2008wt} for recent
overviews). All these processes are among the most promising sources
of gravitational radiation, and gravitational wave observations have
been a key motivation driving numerical relativity.

The exploration of the parameter space of gravitational-wave sources is a
large-scale effort that involves many NR research groups. 
A standard data format is needed to share their results with other research
communities and for collaborative projects. The first example of such an 
application is the use of NR waveforms in gravitational-wave data analysis
codes, but numerical results are also used to produce
analytical template banks, in particular for the case of black-hole
binaries (see e.g.,~\cite{Buonanno:2007pf,Ajith:2007kx}), and in systematic 
comparisons between NR groups~\cite{Hannam:2009hh}.

The aim of this document is to suggest such formats for data exchange
between numerical relativists and the ``numerical relativity data user
community'', in particular gravitational wave data analysts.  It is
clear that there are still outstanding conceptual and numerical issues
remaining in numerical simulations; the goal of this document is not
to resolve them, but to spell out the technical details of the
waveform data in a way that is both precise and sufficiently flexible
to adapt to future research. A primary aim is that NR waveforms can be
incorporated seamlessly within the data-analysis software currently
being developed within the LIGO/Virgo Collaboration (LVC). The
relevant software development is being carried out as part of the LSC
Algorithms Library\footnote{Available from
  \texttt{http://www.lsc-group.phys.uwm.edu/daswg/projects/lal.html}.}
which contains core routines for gravitational-wave data analysis
written in ANSI C99, and is distributed under the GNU General Public
License. 

While it is, in principle, straightforward to extend the data format
to all kinds of gravitational wave sources or NR problems, we first
focus on black-hole-binaries, and discuss in particular applications
to the NINJA project \cite{Aylott:2009ya,Aylott:2009tn}.We also specify
an extension of the data format for neutron star binaries and stellar
collapse.

The key ideas of the data format are as follows:
\begin{itemize}
 \item Simulation data (e.g. time series for different gravitational
  wave strain multipoles) are distributed together with a metadata
  file that describes the data set (including information on authors,
  codes used, physical parameters, links to publications, etc.) and
  contains the file names of the actual simulation data.
      
 \item The metadata file is a simple text file that is easy to read
  and edit for humans, and contains {\tt key = value} pairs, organized
  in {\tt sections}.

 \item In this paper we define the basic syntax of the metadata file,
  together with a collection of keys that would typically be included,
  and the specific set of keys required for submissions to the NINJA
  project.

 \item We also define formats for the gravitational wave strain data
  to be processed with LVC software tools. Formats for other type of
  data, say the time evolution of black hole spins, can
  straightforwardly be defined in analogy.
\end{itemize}

The first version of this document was posted on the arXiv in
September 2007 \cite{Brown:2007jx}, and included a simple data format
for representing the gravitational-wave strain as three-column data
sets of $\{t,h_+,h_x\}$, with the time $t$ being given in equally
spaced intervals, together with a simple metadata format that
specified only black hole (BH) parameters (mass ratio and spin), the authors and
the numerical code. This format has been used for data exchange in a
number of subsequent research projects, notably the first NINJA
project. The present format addresses some of the shortcomings noted
during the first NINJA project, and the requirements of the second
NINJA project, which involves extremely long hybrid 
post-Newtonian-plus-numerical-relativity waveforms, and a larger 
number of waveforms,
which have to be processed automatically.  In order to represent very
long waveforms we here define a second, more economical, data format
for gravitational waveforms.  Also, we have revised the metadata
format as necessary to allow representing all relevant scientific
information, and to facilitate the future use of waveforms from sources that
involve matter.  The current version of this document describes both the
old and new formats, described respectively in
Sections~\ref{sec:format1} and \ref{sec:format2}. Note that citations to
this document prior to 2011 are to version one. 

The remainder of this document is structured as follows: section
\ref{sec:multipoles} describes our conventions for decomposing the
gravitational wave data in terms of spherical harmonics, section
\ref{sec:metadata} specifies the format for metadata, and section
\ref{sec:format} specifies data formats for waveform data.  We
conclude with section \ref{sec:applications} discussing the specific
incarnations of data format specifications for the different phases of
the NINJA project.

\section{Multipole expansion of gravitational waves}
\label{sec:multipoles}

The output of a numerical-relativity code is the full spacetime of a
black-hole-binary system. On the other hand, what is required for
gravitational-wave data-analysis purposes is the strain $h(t)$, as
measured by a detector located far away from the source. The quantity
of interest is therefore the gravitational-wave metric perturbation
$h_{ab}$ in the wave zone, where $a$ and $b$ are space-time indices.
We always work in the Transverse Traceless (TT) gauge so that all
information about the metric perturbation is contained in the TT
tensor $h_{ij}$, where $i$ and $j$ are spatial indices. The wave falls
off as $1/r$ where $r$ is the distance from the source:
\begin{equation}
  \label{eq:1}
  h_{ij} = A_{ij}\frac{M}{r} + \mathcal{O}\left(r^{-2}\right)\,.
\end{equation}
Here $A_{ij}$ is a transverse traceless tensor and $M$ is the total
mass of the system; this approximation is, naturally, only valid far
away from the source.

There are different methods for extracting $h_{ij}$ from a numerical
evolution. One common method is to use the complex Weyl tensor
component $\Psi_4$ which is related to the second time derivative of
$h_{ij}$. Another method is to use the Zerilli function which
approximates the spacetime in the wave-zone as a perturbation of a
Schwarzschild spacetime. For our purposes, it is not important how the
wave is extracted, as different NR groups are free to use methods they
find appropriate. The starting point of our analysis is the multipole
moments of $h_{ij}$ and it is important to describe explicitly our
conventions for the multipole decomposition.
The corresponding values of
$\Psi_4$ or the Zerilli function can be described analogously, in particular
also regarding the data formats described later. Since the
strain and $\Psi_4$ or the Zerilli functions are related by numerical operations
and other potential ambiguities, it is often useful to have these original
data available for consistency checks with the strain waveforms.

Let $(x,y,z,t)$ be a Cartesian coordinate system in the wave zone,
sufficiently far away from the source. Let $\xhat$, $\yhat$ and
$\zhat$ denote the coordinate basis vectors.  Given this coordinate
system, we define standard spherical coordinates $(r,\iota,\phi)$
where $\iota$ is the inclination angle from the $z$-axis and $\phi$ is
the phase angle. At this point, we have not specified anything about
the source. In fact, the source could be a binary system, a star
undergoing gravitational collapse or anything else that could be of
interest for gravitational wave source modeling.  In later sections
we will specialize to particular GW sources and suggest
possibilities for some of the various choices that have to be
made. However, as far as possible, these choices are eventually to be
made by the individual source modeling group.

We break up $h_{ij}$ into modes in this coordinate system. In the wave
zone, the wave will be propagating in the direction of the radial unit
vector
\begin{subequations}
  \begin{alignat}{3}
    \rhat &= &\xhat&\,\sin\iota\cos\phi\, &+&\,
    \yhat\,\sin\iota\sin\phi + \zhat\,\cos\iota\,.  \intertext{A
      natural set of orthogonal basis vectors from which to build the
      transverse-traceless basis tensors is} \iotahat &=
    &\xhat&\,\cos\iota\cos\phi \,&+&\, \yhat\,\cos\iota\sin\phi
    - \zhat\,\sin\iota \,,\\
    \phihat &= -\,&\xhat&\,\sin\phi \,&+&\, \yhat\,\cos\phi\,.
  \end{alignat}
\end{subequations}
In the transverse traceless gauge, $h_{ij}$ has two independent
polarizations
\begin{equation}
  \label{eq:2}
  \tens{h} = \sum_{i,j}h_{ij}\,\ihat\otimes\jhat = h_+ \eplus +
  h_\times \ecross\,,
\end{equation}
where $\eplus$ and $\ecross$ are the usual basis tensors for
transverse-traceless tensors in the wave frame
\begin{equation}
  \label{eq:9}
  \eplus = \iotahat\otimes\iotahat - \phihat\otimes\phihat\,, \qquad
  \textrm{and} \qquad \ecross = \iotahat\otimes\phihat +
  \phihat\otimes\iotahat\,.
\end{equation}
It is convenient to use the combination $h_+ - ih_\times$, which is
related to $\Psi_4$ by two time derivatives\footnote{We define
  $\Psi_4$ as $\Psi_4 \define C_{abcd}\bar{m}^a n^b \bar{m}^c n^d$
  where $C_{abcd}$ is the Weyl tensor and $a,b\ldots$ denote abstract
  spacetime indices. If we denote the unit timelike normal to the
  spatial slice as $e_{\hat{t}}^a$ and the promotions of
  $\{\rhat,\iotahat,\phihat\}$ to the full spacetime as
  $\{e_{\hat{r}}^a,e_{\hat{\iota}}^a,e_{\hat{\phi}}^a\}$, then the
  null tetrad adapted to the constant $r$ spheres is
  $\{\ell^a,n^a,m^a,\bar{m}^a\}$ where $\ell^a = (e_{\hat{t}}^a +
  e_{\hat{r}}^a)/\sqrt{2}$, $n^a = (e_{\hat{t}}^a -
  e_{\hat{r}}^a)/\sqrt{2}$, $m^a = (e_{\hat{\iota}}^a +
  ie_{\hat{\phi}}^a)/\sqrt{2}$, and $\bar{m}^a$ is the complex
  conjugate of $m^a$.}
\begin{equation}
  \label{eq:3}
  \Psi_4 = \ddot{h}_+ - i\ddot{h}_\times\,.
\end{equation}
It can be shown that $h_+-ih_\times$ can be decomposed into modes
using spin-weighted spherical harmonics $\Ys_{lm}$ of weight $-2$:
\begin{equation}
  \label{eq:hDecomposition}
  h_+ - ih_\times =
  \frac{M}{r}\sum_{\ell=2}^{\infty}\sum_{m=-\ell}^\ell H_{\ell m}(t)\,
  \Ytwo_{\ell m}(\iota,\phi)\,.
\end{equation}
The expansion parameters $H_{lm}$ are complex functions of the
retarded time $t-r$ and, if we fix $r$ to be the radius of the sphere
at which we extract waves, then $H_{lm}$ are functions of $t$ only.

The explicit expression for the spin-weighted spherical harmonics in
terms of the Wigner $d$-functions is
\begin{equation}
  \label{eq:5}
  \Ys_{lm} = (-1)^s\sqrt{\frac{2\ell+1}{4\pi}}
  d^{\,\ell}_{m,s}(\iota)e^{im\phi},
\end{equation}
where
\begin{equation}
  \label{eq:6}
  d^{\,\ell}_{m,s}(\iota) = \sum_{k = k_1}^{k_2}
  \frac{(-1)^k[(\ell+m)!(\ell-m)!(\ell+s)!(\ell-s)!]^{1/2}}{(\ell +m
    -k)!(\ell-s-k)!k!(k+s-m)!}  \times
  \left(\cos\left(\frac{\iota}{2}\right)\right)^{2\ell+m-s-2k}\left(\sin\left(\frac{\iota}{2}\right)\right)^{2k+s-m}
\end{equation}
with $k_1 = \textrm{max}(0, m-s)$ and $k_2=\textrm{min}(\ell+m,
\ell-s)$. For reference,
\begin{eqnarray}
  \label{eq:7}
  \Ytwo_{22} &=& \sqrt{\frac{5}{64\pi}}(1+\cos\iota)^2e^{2i\phi} \,,\\
  \Ytwo_{21} &=& \sqrt{\frac{5}{16\pi}}  \sin\iota( 1 + \cos\iota )e^{i\phi} \,,\\
  \Ytwo_{20} &=& \sqrt{\frac{15}{32\pi}} \sin^2\iota \,,\\
  \Ytwo_{2-1} &=& \sqrt{\frac{5}{16\pi}}  \sin\iota( 1 - \cos\iota
  )e^{-i\phi} \,,\\
  \Ytwo_{2-2} &=& \sqrt{\frac{5}{64\pi}}(1-\cos\iota)^2e^{-2i\phi}\,.
\end{eqnarray}
The mode expansion coefficients $H_{lm}$ are given by
\begin{equation}
  \label{eq:10}
  H_{\ell m} = \frac{1}{M}\, \oint \Ytwo_{lm}^\star
  (\iota,\phi)(rh_+-irh_\times) \,d\Omega\,.
\end{equation}
If $\Psi_4$ is used for wave extraction, then $H_{lm}$ is given by two
time integrals of the corresponding mode of $\Psi_4$. In this case, it
is important that the information provided contains details about how
the integration constants are chosen. We define $h_+^{(\ell m)}$ and
$h_\times^{(\ell m)}$ as the real and imaginary parts of the $H_{lm}$
according to
\begin{equation}
  \label{eq:11}
  rh_+^{(\ell m)}(t) -irh_\times^{(\ell m)}(t) \define M\,H_{\ell
    m}(t)\,.
\end{equation}
It is these modes $rh_{+,\times}^{(\ell m)}$ of $rh_+$ and $rh_\times$
that we suggest to be provided as functions of time in units of $M$
for vacuum spacetimes and in units of $M= 1 M_\odot \hat{=}\,
4.92549059 \times 10^{-6}\,\mathrm{s}$ for spacetimes involving matter
(using the values of $M_\odot$ and fundamental constants listed in
section~E.3 of \cite{lalspec}).

\subsection{Application to Waveforms extracted in the Quadrupole
  Approximation}

Many simulation codes for non-vacuum spacetimes still operate in
Newtonian gravity and dynamics or employ post-Newtonian or
conformally-flat approximations to general relativity. Such codes
generally follow the quadrupole approximation for estimating
gravitational waveforms. In this approximation, the
transverse-traceless gravitational wave field is related to the second
time derivative of the reduced mass quadrupole tensor
$\ddot{I}\barie_{ij}$ as
\begin{equation}
  r h_{ij}^{TT,\mathrm{quad}} = \frac{2 G}{c^4} \Pi_{ijkl}
  \ddot{I}\barie_{kl}\,\,,
\end{equation}
where $\Pi_{ijkl}$ is the necessary projection operator to transform
the right-hand side into the transverse-traceless gauge. Here and in the
following we assume that $\ddot{I}\barie_{ij}$ is given in standard
cgs units. $\ddot{I}\barie_{ij}$ is a common wave extraction output of
simulation codes. In order to express $H_{2m}$ in terms of
$\ddot{I}\barie_{ij}$, one first expresses $h_+(\iota,\phi)$ and
$h_\times(\iota,\phi)$ in terms of $\ddot{I}\barie_{kl}$, then
convolves these with $^{-2}Y^*_{lm}$ (cf. eq.~\ref{eq:10}). The result
is
\begin{eqnarray}
  H^\mathrm{quad}_{20}    &=& \sqrt{\frac{32\pi}{15}} \frac{G}{c^4} 
  \left(\ddot{I}\barie_{zz} 
    - \frac{1}{2}(\ddot{I}\barie_{xx}
    +\ddot{I}\barie_{yy})\right)\,\,,\\
  H^\mathrm{quad}_{2\pm1} &=& \sqrt{\frac{16\pi}{5}} \frac{G}{c^4}
  \left(\mp \ddot{I}\barie_{xz} 
    + i \ddot{I}\barie_{yz}\right)\,\,,\\
  H^\mathrm{quad}_{2\pm2} &=& \sqrt{\frac{4\pi}{5}} \frac{G}{c^4} 
  \left(\ddot{I}\barie_{xx} 
    - \ddot{I}\barie_{yy} \mp 
    2 i \ddot{I}\barie_{xy} \right)\,\,.
\end{eqnarray}

\section{Metadata format}\label{sec:metadata}

Simulation data are distributed together with a metadata file that
describes the data set and contains the file names of the actual
simulation data.  The metadata file is a simple text file that is easy
to read and edit for humans, and contains {\tt key = value} pairs,
organized in {\tt sections}, adopting the format that is very commonly
used for software configuration files.  The metadata file aims in
particular to allow automated production of waveform tables such as
the simulation tables in the NINJA and Samurai
papers~\cite{Aylott:2009ya,Aylott:2009tn,Hannam:2009hh}, and to
provide all the scientific information that is required to process the
data sets.

This metadata file contains (at least) two sections, one for the
simulation metadata, and the other listing the filenames which
correspond to the various $(\ell,m)$ modes of the waveform. A separate
metadata file is required for each simulation.

The metadata format proposed here is based on that used for the first
NINJA project, and described in detail in Version 1 of this document
\cite{Brown:2007jx}, which included a minimum essential set of
information related to a given numerical simulation---the NR group
that produced the data, their contact information, and the initial
spins of the two black holes, their mass ratio, and the initial
frequency of the waveform.

During the course of the first NINJA project, and discussions related
to follow-up projects, it became clear that the metadata format needs
to be extended to include more information on the source of the
waveforms, their physical properties, their accuracy, the methods used
to combine them with PN/EOB results to produce hybrid waveforms, and
the availability of these waveforms for use by others in scientific
projects. A more precisely defined metadata format is also required to
allow the automated construction of waveform catalogs.

Examples of the metadata format are given at the end of this
document. Further details, and modifications more recent than the
production of this document, can be found on the NINJA project
website, {\tt www.ninja-project.org}.

\subsection{Syntax for sections, entries and comments}

A metadata file consists of sections headed by
\begin{verbatim}
[sectionname]
\end{verbatim} and entries of the form 
\begin{verbatim}
key = value
\end{verbatim}

The {\tt key} strings, which we will also refer to as tags,
are not allowed to contain spaces, but spaces can be used
in the value field, where they are interpreted as field separators or
text depending on the context implied by {\tt key}.

Values default to those set in the {\tt[metadata]} section, and values
set in any other section override the values set in the {\tt
  [metadata]} section, and are defined only locally in that section.

Keys that point to data files (such as mode names, i.e ``2,-2'', see
below) are ignored in the {\tt [metadata]} section. All keys that can
appear in a {\tt [metadata]} section can also appear in other
sections.

The comment character is ``\texttt{\#}''---all text to the right of a comment
character is interpreted as a comment.

\subsection{Format of 'names'}

Names of files (metadata files, simulation data, documentation files,
etc.)  and {\tt-name} keys, such as {\tt simulation-name} should {\it
  only} contain alphanumeric characters plus ``\texttt{-}'', ``\texttt{\_}'', ``\texttt{.}''.

The suggested naming convention for binary black hole metadata files
is ``name.bbh''.

\subsection{Format of email addresses}

To indicate the full name of authors and waveform submitters, the
format Joan Smith $<$joan.smith@example.edu$>$ should be used
consistently, i.e. the email address field should contain the real name.

\subsection{Documentation and References}

Any tag can be extended by the following tags as tag-subtag to add
documentation or references:

\begin{verbatim}
bibtex-keys   #  referring to a SPIRES or ADS style bibtex key, SPIRES 
              #  keys can easily be expanded to complete bibtex entries.
comments      #  to add a brief comment text.
documentation #  to add pointers to further documentation, i.e. file names or URLs.
\end{verbatim}

For example, the value obtained for the eccentricity of a binary could
be documented as
\begin{verbatim}
eccentricity = 0.001
bibtex-keys    = Einstein:1905xx
comments       = our method for this case differs from our other submissions!
documentation  = eccentricity_writeup.tex # file included with the submission. 
\end{verbatim}

\subsection{Units}
\label{sec:units}

Black hole waveforms are given in units of $M$, where $M = m_1+m_2$ is
the total initial black hole mass (for spinning BHs, this mass
includes the spin contribution).  Black hole separations are measured
in coordinate distance, measured in $M$.

For simulations involving matter fields, we use the convention of time
in units of $M_\odot=4.92549095\times 10^{-6}\,\mathrm{s}$, and strain
$rh_{\{+,\times\}}$ in units of
$M_\odot=1.47662504\times10^{5}\,\mathrm{cm}$. Frequencies in metadata
are in units of $1/M_\odot= 203.025447\, \mathrm{kHz}$. These geometrized solar
units are compatible with LAL values of $M_\odot$ and fundamental
constants listed in section~E.3 of \cite{lalspec}. Use of these units
can be flagged with metadata \texttt{mass-scale = 1}.

\subsection{Gravitational wave start frequency}

For use in gravitational wave data analysis, it is essential to
indicate the start frequency of the waveform. Since different spheroidal
harmonic modes differ in frequency, we choose to specify the start
frequency of the $l=m=2$ harmonic with the key {\tt freq-start-22}.
For black hole binaries, where all results scale with the total mass $M$, 
this is best done in units of $M$, and frequency $f_0/M$ is specified.
For a given value of the mass, this gives the physical start frequency
of the waveform, and this will need to be less than the lower cut-off
frequency relevant for a particular detector.  For example, $f_0 =
40\,$Hz is an appropriate value for the initial LIGO detectors, and
$f_0 = 10\,$Hz is appropriate for advanced detectors.

\subsection{Accuracy measures and error bars}

Any number-valued tag also implies the tags {\tt tag-error-relative}
and {\tt tag-error-range} to quote relative and absolute errors.  We
use the following syntax:
\begin{verbatim}
value-error-relative = 0.01
\end{verbatim}
means tag has a relative error of 1\%, while
\begin{verbatim}
value-error-range = 3 9
\end{verbatim}
means $3 \leq {\rm value} \leq 9$.

Useful accuracy measures on time series data are more difficult to
specify, and depend on the application. For waveform data,
mismatch values can be defined. For the NINJA project, we have
specified three tags that relate to the accuracy of the
waveforms---{\tt uncertainty-in-number-of-cyles}, {\tt
  relative-amplitude-uncertainty}, and {\tt mismatch}.  Details of how
these uncertainties are defined (in particular for what follows the
tag ``mismatch'') can be specified in comments fields or decided
upon for specific projects. These three quantities
are relatively easy to calculate, and they can be easily extended to more 
sophisticated accuracy measures in the future.

\subsection{Vector valued tags}

For vector-valued tags we use the syntax
\begin{verbatim}
value = {1, 2, 3}.
\end{verbatim}
Relative and absolute errors can be specified as
\begin{verbatim}
value-error-relative = {.01, .01, .01}
\end{verbatim}
or
\begin{verbatim}
value-error-range = {.5, 1.5, 2.5} {4, 4, 4}.
\end{verbatim}

\subsection{Black-hole binaries}

\subsubsection{Parameterization of parameter space}

A numerical-relativity simulation of  black-hole
binary coalescence  has many parameters that need to be specified, 
and several of them may not be directly relevant to the data-analysis problem.
We need to determine which parameters of the numerical simulation are useful
for the users of the data.  For our purposes, a single numerical waveform is
defined by at least seven parameters: the mass ratio $q = M_1/M_2$ and the
three components of the individual spins $\vec{S}_1$ and
$\vec{S}_2$. The choice of precisely how $M_1$, $M_2$ and the
spins are calculated is left up to the individual numerical relativity
groups. Note that for systems with matter, as discussed below, it makes more
sense to specify all masses in physical units, e.g. in solar masses, instead
of using the mass ratio.

The mass ratio $q$ is assumed to be in the form $q = m_1 /m_2$.
It is essential to adhere to this definition for spinning binaries, because this is the only way for
an automated script to determine which spin belongs to which black hole.
In addition, for simplicity we suggest to adhere to the convention $m_1 \geq m_2$,
i.e., the first BH is the heavier one.

\subsubsection{Coordinate system}

In the default coordinate system, all binaries start out in the
$xy$-plane, consistent with the convention of Table 1 in the first
NINJA paper~\cite{Aylott:2009ya}.

For aligned-spin systems, it is recommended that the initial momenta
be tangent to the $xy$-plane. For precessing systems this will not
necessarily be convenient, and BH momenta or velocity vectors should
be specified. To fix the rotation in the $xy$-plane, the key {\tt
  initial-separation-angle} specifies the direction of the separation
vector from the first to the second BH in degrees, e.g., if this
direction points along the $x$-axis then
\begin{verbatim}
initial-separation-angle = 0
\end{verbatim}
and if this direction points along the negative $y$-axis then
\begin{verbatim}
initial-separation-angle = 270  #  or initial-separation-angle = -90.
\end{verbatim}
Specifying a value for {\tt initial-separation-angle} implies
\begin{verbatim}
 initial-bh-position1z = 0
 initial-bh-position2z = 0
\end{verbatim}
and sets {\tt initial-bh-position1x}, {\tt initial-bh-position1y},
{\tt initial-bh-position2x}, {\tt initial-bh-position2y} accordingly,
assuming a center of mass frame and using the value for {\tt
  mass-ratio}.

In order to override these definitions of the coordinates, specify the
keys
\begin{verbatim}
initial-bh-position1x, initial-bh-position1y, initial-bh-position1z.
\end{verbatim}
In order to specify the initial motion of the BHs, define the momenta
or velocities,
\begin{verbatim}
initial-bh-momentum1x, initial-bh-momentum1y, initial-bh-momentum1z, 
initial-bh-velocity1x, initial-bh-velocity1y, initial-bh-velocity1z, 
\end{verbatim}
and analogously for the second BH. If one component of the momenta or
velocities is specified, the other components default to zero.

\subsection{Binaries involving neutron stars}

Double-neutron-star or mixed black-hole/neutron-star binaries have
many characteristics in common with binary-black-hole systems. When
the bodies are far apart, the dynamics are also determined by total
mass, mass ratio, and individual spins, as in the specification above.

However, the properties of matter set a physical scale which fixes,
for example, the radius of a neutron star relative to the length scale
of the mass. Waveforms from such systems cannot be scaled to
different masses, and  the masses of all objects have to be specified
explicitly using fixed units of $M_\odot$ as discussed in Sec.~\ref{sec:units}. 
Thus, unlike in the simpler BBH case, the mass ratio is not used to parameterize
configurations. Note that the convention $q \leq 1$ in the neutron star
simulation community is opposite to the BBH data format convention.

In the simplest case the equation of state (EOS) for a simulated
neutron star may be a polytrope or ideal fluid characterized by two
parameters of adiabatic index and compactness; more generally a cold
EOS is an arbitrary tabled or parameterized function of pressure
vs. density, and even more generally details of heating and
composition may also be described (and influence system
dynamics). Simple metadata descriptions are therefore more
challenging, and full information will most likely need to be given in
comments or \texttt{simulation-details}. If useful characteristic
parameters can be agreed on (for example compactness, tidal
deformability, tidal disruption frequency estimates), these can be included in the metadata.

Since systems involving neutron stars (except for extreme mass ratio
cases) are low mass systems, hybrid waveform construction will be
required for most realistic detector injections.

\subsection{Stellar Collapse and Single Compact-Star Spacetimes}

Simulations of stellar collapse, core-collapse supernovae, and single
compact stars (neutron stars, strange stars, etc.) have less global
parameters, but the physics involved in their modeling is generally
more complex, is frequently done under some symmetry assumptions and
involves complicated physics, such as nuclear equations of state,
neutrino transport, realistic initial stellar models, etc. Depending on
application, this modeling complexity is reflected in a more
complicated and extensive set of metadata information needed to
characterize a given waveform. An example is given in section
\ref{sec:ninjamatter}.

A general convention to follow is that the system spin should be
aligned with the positive $z$-axis.

\section{Data formats for waveforms}
\label{sec:format}

In the following subsections we outline two data formats for
waveforms, represented as multipole components of the complex wave
strain $h(t)$.

The first format directly represents the strain $h(t)$, and the second
instead specifies the phase and amplitude of the wave, which can then
be used to reconstruct the wave strain. The second format has the
advantage that the phase and amplitude functions factor out the
oscillations of $h(t)$ with the gravitational wave frequency, and can
thus be accurately represented with fewer data points.

Data in both formats are written as 3-column ASCII plain text files,
which can optionally be compressed with gzip (i.e., file readers for
such data are supposed to support gzip).
Note that time steps need not be equidistant, which was required in
the first version of the data format~\cite{Brown:2007jx}.

For a given simulation, numerical groups decide the
maximum value of $\ell = \ell_{\rm max}$ to which they will provide
the waveform. For every $\ell \leq \ell_{\rm max}$, waveforms must be 
provided for all values of $m = -\ell, \ldots, \ell$, irrespective of any
symmetries that may be present in the simulation. If there are certain
modes which vanish, or have a small amplitude which cannot be accurately determined,
then such data can either be set to zero, or equivalently left out.
Thus, any modes which will not be provided will be assumed to 
have zero amplitude, and providing an $(\ell,m)$-mode without
the corresponding $(\ell,-m)$-mode will in general lead to
incorrect results.

It is natural to use the total mass $M$ of the binary as the unit for
the time and strain columns. However, there are subtleties in the
choice of $M$, and in the definition of black hole masses in a
dynamical spacetime, e.g. one could alternatively use approximations
to the ADM or Bondi mass of the spacetime. Similar
ambiguities exist for other relevant quantities, such as the black
hole spins. While for many applications these subtleties have little
or no relevance, it is important that all definitions can be
reproduced, and therefore the metadata file should include
links to sufficient documentation on the details for that simulation.

For simplicity, we identify $M$ with an estimate of the
total initial black hole mass, and clearly highlight deviations from
this convention.

\subsection{Waveforms represented as a complex time series}
\label{sec:format1}

In this format, the data for a single mode $rh_{+,\times}^{(\ell m)}$
is written as a plain text file in three columns for the time $t$,
$rh_+^{(\ell m)}$ and $rh_\times^{(\ell m)}$ respectively.

The strain multiplied by the distance will also be in units of the
total mass $M$ of the binary. There can be any number of comment lines
at the top of the file; it should however be noted that this
can put restrictions on the range of plotting tools that can be used. 
It is recommended that 
all information contained in such comments is also available in the
metadata files. 

For uniformly sampled waveforms, a rate of $1 \times M / m$ (where $m$
labels the harmonic) is believed to be sufficient for most data
analysis purposes, and is recommended unless a more careful analysis
suggests the use of a different value for a particular case. This time
resolution is generally sufficient during the merger and ringdown,
when the waveform frequency is highest; it is acceptable to use a
coarser time sampling at lower frequencies.

Gravitational strain waveforms in the complex time series
representation are referenced in a \texttt{ht-data} section of the
metadata file, e.g.~as
\begin{verbatim}
[ht-data]
2,2  = hmod.r5.l3.l2.m2 
2,1  = hmod.r5.l3.l2.m1
2,-1 = hmod.r5.l3.l2.m-1
2,-2 = hmod.r5.l3.l2.m-2.
\end{verbatim}
For $\Psi_4$ or Zerilli waveforms, the section headers should 
be \texttt{[psi4t-data]} and  \texttt{[zerillit-data]}.

An example data file looks like:

\begin{verbatim}
# numerical waveform from .... 
# equal mass, non spinning, 5 orbits, l=m=2
# time       hplus        hcross
0.000000e+00 1.138725e-02 -8.319811e-04
2.000000e-01 1.138725e-02 -1.247969e-03
4.000000e-01 1.138726e-02 -1.663954e-03
6.000000e-01 1.138727e-02 -2.079936e-03
8.000000e-01 1.138728e-02 -2.495913e-03
1.000000e-00 1.138728e-02 -2.911884e-03
1.200000e+00 1.138729e-02 -3.327850e-03
1.400000e+00 1.138730e-02 -3.743807e-03
1.600000e+00 1.138731e-02 -4.159757e-03
1.800000e+00 1.138733e-02 -4.575696e-03
2.000000e+00 1.138734e-02 -4.991627e-03
2.200000e+00 1.138735e-02 -5.407545e-03
2.400000e+00 1.138737e-02 -5.823452e-03
2.600000e+00 1.138739e-02 -6.239345e-03
2.800000e+00 1.138740e-02 -6.655225e-03
3.000000e+00 1.138752e-02 -7.071059e-03
3.200000e+00 1.138754e-02 -7.486903e-03
3.400000e+00 1.138757e-02 -7.902739e-03
......
\end{verbatim}

\subsection{Waveforms represented by phase and amplitude}
\label{sec:format2}

NINJA waveforms need to be scaled across a wide range of masses, and
injected without aliasing into the data stream.  This presents two
conflicting motivations.  First, in order to cover the low frequency
bands of advanced detectors, simulating low-mass systems, the
waveforms need to be very long.  On the other hand, in order to
prevent aliasing---especially of the merger and ringdown of high-mass
systems---the data needs to be sampled very finely.  Naively, these
two problems taken together would lead to enormous data sets,
especially as NINJA extends to more interesting regimes of
low-frequency sensitivities and low-mass systems.  By changing the
format of stored data, we can sample the waveform very coarsely during
the long inspiral, and very finely during the highly dynamic merger
and ringdown, then reconstitute the data as necessary.

We have chosen to represent the waveform through the wave phase and
amplitude, because they are typically very simple functions of time,
meaning that interpolation is quite accurate, so the waveforms can be
represented with a relatively coarse time sampling.  In this format,
the data for the phase and amplitude of a single mode of
$rh_{+,\times}^{\ell m}$ is written as a plain text file in three
columns for the time $t$, amplitude $A$, and phase $\phi$,
respectively. The wave strain at a given time can be reconstructed by
\begin{equation}
  r h^{\ell m}_+ - i rh^{\ell m}_\times = A_{\ell m} e^{i \phi_{\ell
      m}}~. \label{eq:strain}
\end{equation}
Because the strain is insensitive to offsets of $2\pi$ in the phase, a
stepwise function of integer multiples may be added to $\phi_{\ell m}$
to make it somewhat continuous; the objective is for interpolation of
the time series for $\phi_{\ell m}$ to be accurate.  With these
choices of variables, the time sampling need not be uniform.  

The only requirement is that the time sampling be fine enough at any
given time that the wave strain can be reconstructed to be identical
(within some desired accuracy) to the ``original'' waveform phase and
amplitude data using only first-order interpolation between the data
points provided. Applying higher-order interpolation to the provided
data will presumably yield results with greater fidelity to the
original hybrid waveform, and may be useful if it is expected that
accurate derivatives will need to be taken of the waveform.  However,
the defining requirement is that the phase and amplitude can be
accurately reconstructed using only first-order interpolation.  The
accuracy of the reconstruction is left to contributing NR groups.  In
general, two separate tolerances will need to be defined:
$\text{Tol}_{A}$ and $\text{Tol}_{\phi}$.  Presumably, these will be
chosen to be roughly the same as the estimated accuracy of the input
data.

Within this requirement, NR groups may choose any method to sample the
data points. A simple but robust algorithm for removing unnecessary
time steps from the data is as follows.  First, we assume that both
the initial and final time steps should be included in the final data
set.  Next is a recursive stage in which each interval of the new
coarser time series is checked to ensure that $\phi_{\ell m}$ at the
midpoint of that interval can be linearly interpolated to within the
desired tolerance of that point in the original data set.  If the
interpolation is not sufficiently accurate, the midpoint is included
in the coarse set, and the test continues on the two new intervals
thus formed.  Finally, another recursive stage checks each point of
the input data set to ensure that both $A_{\ell m}$ and $\phi_{\ell
  m}$ can be correctly reproduced.  If not, we include the midpoint of
the coarse interval in which that data point is found, and continue
and repeat the check.

This algorithm has been implemented in the NINJA code repository as
\texttt{MinimizeGrid}.  Depending on the length of the waveform, it
reduces the size of the data set by anywhere from a few percent for
short numerical data to 99\% for very long hybrid waveforms.  
For the 
long waveforms needed in the second NINJA project (to allow injection
into detector noise at low masses),  this technique
is crucial to make the data sets manageable.

Gravitational-strain waveforms in the phase-amplitude series
representation are referenced in a \texttt{ht-phiamp-data} (alternatively
 \texttt{psi4t-data} and  \texttt{zerillit-data})
section of
the metadata file as, e.g.,
\begin{verbatim}
[ht-phiamp-data]
2,2  = hmod.r5.l3.l2.m2
2,1  = hmod.r5.l3.l2.m1
2,-1 = hmod.r5.l3.l2.m-1
2,-2 = hmod.r5.l3.l2.m-2.
\end{verbatim}

\section{Applications}\label{sec:applications}

In this section we list the compulsory metadata fields for some
concrete projects, in particular the first NINJA project, the NINJA
waveform catalog---a publicly available catalog of BBH simulation
metadata from simulations that have been performed in the NR
community---, the second NINJA project, and the Matter NINJA proposal.

\subsection{First NINJA project}

This is an example for the old format \cite{Brown:2007jx}, used in the
first NINJA project \cite{Aylott:2009ya,Aylott:2009tn}

\begin{verbatim}
[metadata]
simulation-details = initial separation 11M, QC parameters
nr-group = friendlynrgroup 
email = sub.mitter@good-science.org
mass-ratio = 1.0
spin1x = 0.0
spin1y = 0.0
spin1z = 0.5
spin2x = 0.0
spin2y = 0.0
spin2z = 0.5
freqStart22 = 0.05

[ht-data]
2,2  = hmod.r5.l3.l2.m2 
2,1  = hmod.r5.l3.l2.m1
2,-1 = hmod.r5.l3.l2.m-1
2,-2 = hmod.r5.l3.l2.m-2.
\end{verbatim}

\subsection{NINJA waveform catalog}

A waveform catalog is currently under construction at {\tt
  www.ninja-project.org}, for use in the second and ongoing NINJA
projects. For this catalog, the compulsory fields are give below,
and examples of options are given in Sec.~\ref{sec:ninja2}.

\begin{verbatim}
[metadata] section  

license = ninja2
documentation
publication             # can be  "none"
simulation-bibtex-keys
simulation-name
simulation-uuid
authors-tag
submitter-email
authors-emails
nr-uuid,data-type = hybrid
code
code-bibtex-keys
initial-separation
initial-data-type
initial-data-bibtex-keys
quasicircular-bibtex-keys
eccentricity
mass-ratio
spin1x
spin1y
spin1z
spin2x
spin2y
spin2z
freq-start-22
number-of-cycles-22
extraction-radius
uncertainty-in-number-of-cycles
relative-amplitude-uncertainty 
mismatch

[ht-data] section   

Entries for the 2,2 and 2,-2 modes.
\end{verbatim}

\subsection{NINJA 2}
\label{sec:ninja2}
This is a suggested format for the Ninja2 black holes project, \\
\url{http://www.ninja-project.org/doku.php?id=ninja2:home}. 
Slashes are used to indicate suggested options.

\begin{verbatim}
[metadata]
license                = ninja-catalog/ninja2/thirdparty/private/public
comments               = 
documentation          = short.txt  myfile.pdf 
publication            = arXiv:0901.4399/unpublished/inprint
simulation-bibtex-keys = Aylott:2009ya

simulation-name = spp50 
simulation-uuid = 373b1340-db3d-11de-aef4-0002a5d5c51b
authors-tag     = SubCorrIts 
submitter-email = Sub Mitter <submitting@author.edu>
authors-emails  = Corr Espondent<corresponding@author.edu>,  Its Me <somebody@else.edu>

data-type = NR/PN/hybrid
hybrid-nr-uuid = 373b1340-db3d-11de-aef4-0002a5d5c51ba  # optional
hybrid-approximant = TaylorT1                           # only for hybrids
hybrid-method = hybrid-doc.txt                          # only for hybrids

simulation-relative-resolution = 3d18fda3-c58d-4a95-9a87-b1275da258e6   2 
# resolution is 2 times better than for this simulation-uuid

code             = mycode
code-version     = 247
code-bibtex-keys =

initial-data-type  = Bowen-York quasicircular 
initial-data-uuid =
initial-data-bibtex-keys = 
quasicircular-bibtex-keys =

initial-separation = 11
initial-separation-angle = 90 # only for aligned spins
                              # alternatively specify initial-bh-position1 etc. 

# direction of the separation vector from the first to the second BH, in degrees 
initial-bh-momentum1x = -0.0900993   
# if momenta for second BH are not given, we are in the center of mass frame
initial-bh-momentum1y = -0.000709412 
# alternatively, use initial-bh-velocity1x etc.
initial-bh-momentum1z = 0

eccentricity = 0.002 0.001
eccentricity-error-range = 0 0.003
mass-ratio = 1.0

spin1x = 0.0
spin1y = 0.0
spin1z = 0.5
spin2x = 0.0
spin2y = 0.0
spin2z = 0.5

freq-start-22/freqStart22 = 0.05
number-of-cycles-22 = 19.3

extraction-radius = somenumber/infinity/extrapolated
extraction-techniques =
number-of-cycles-22-error-relative = 0.001
amplitude-error-relative = 0.02

[ht-data]
2,2  = hmod.r5.l3.l2.m2 
2,1  = hmod.r5.l3.l2.m1
2,-1 = hmod.r5.l3.l2.m-1
2,-2 = hmod.r5.l3.l2.m-2.
\end{verbatim}

\subsection{First Matter NINJA project}
\label{sec:ninjamatter}
This is a suggested format for a project such as
\url{http://www.ninja-project.org/doku.php?id=matter:home}. Slashes
are used to indicate suggested options.
\subsubsection{Binary Systems}
\begin{verbatim}
[metadata]
simulation-type = BNS/NSBH
mass-scale = 1

documentation = short.txt
publication = arXiv:1001.1234/unpublished/inprint

submitter-email = Sub Mitter <submitting@author.edu>
authors-email = Cor Espondent<corresponding@author.edu>, Sub Mitter <submitting@author.edu> 

#Use convention that body 1 is BH in NSBH
#Use convention that body 1 is larger NS in NSNS
mass1 = 1.4
mass2 = 1.4
# calculate mass-ratio from mass specification as needed
# to avoid issues of convention conflict
spin1x = 0.0
spin1y = 0.0
spin1z = 0.0
spin2x = 0.0
spin2y = 0.0
spin2z = 0.0

# simulation setup and physics 
gravity-type = NR/CFC/ApproxGR/Newtonian
fluid-type   = GRHD/GRMHD/HD/MHD/OtherType
symmetries   = 3Dbitant/3Dnone/OtherSymmetry
eos-name = Gamma2Poly/APR/LS/PP-HB/OtherName
eos-reference = SPIRES bibtex code or ADS link
eos-details = Specify EOS parameters (string may be long)
eos-details-reference = SPIRES bibtex code or ADS link
neutrino-treatment = None/CoolingFunction/Leakage/MGFLD/OtherName
neutrino-treatment-reference = SPIRES bibtex code or ADS link

# waveform properties
inspiral = Y/N							# suitable target for inspiral search?
freq-Start-22 = 1.50E-4			# in 1/M_sun = 203.025447 kHz
data-type = NR/PN/hybrid
# pn-method = pn-doc.txt
# hybrid-method = hybrid-doc.txt 

[ht-data]
2,2  = example1_l2_m2.dat
2,-2 = example1_l2_m-2.dat.
\end{verbatim}

\subsubsection{Single-star systems}
\begin{verbatim}
[metadata]
simulation-type = CCSN/NSCollapse/NSOscillations/OtherType
mass-scale = 1

documentation = short.txt
publication = SPIRES or ADS link/arXiv:1001.1234/unpublished/inprint

submitter-email = Sub Mitter <submitting@author.edu>
authors-email = Cor Espondent <corresponding@author.edu>, Sub Mitter <submitting@author.edu> 

mass = 2.8 #total system gravitational mass in solar masses (M_Sun)
mass-baryonic = 3.0 #total system gravitational mass in M_Sun
spin = 1.0 #total system spin in c=G=M_Sun=1

# simulation setup and physics 
gravity-type = NR/CFC/ApproxGR/Newtonian
fluid-type   = GRHD/GRMHD/HD/MHD/OtherType
symmetries   = axi/3Doctant/3Dbitant/3Dquadrant/3Dnone

# rotation
rotation   = describe how rotation is set up; define meaning of rotation parameters
rotation-parameters = specify rotation parameters (string may be long)
rotation-reference = SPIRES bibtex code or ADS link

# microphysics
eos-name = Hybrid/HShen/LS/OtherName
eos-reference = SPIRES bibtex code or ADS link
eos-details = Specify EOS parameters (string may be long)
eos-details-reference = SPIRES bibtex code or ADS link

neutrino-treatment = None/CoolingFunction/Leakage/MGFLD/OtherName
neutrino-treatment-reference = SPIRES bibtex code or ADS link

# describe waveform type
data-type  = NR/Quadrupole/OtherName
inspiral = N

freq-Start-20 = 0.0
freq-Start-22 = 0.0

[ht-data]
2,0  = example1_20.dat
2,2  = example1_22.dat.
\end{verbatim}

\section*{Acknowledgments}

We are grateful to various numerical relativists for numerous
discussions and suggestions. In particular, we would like to thank the
following people for valuable inputs to this document: Peter Diener,
Luis Lehner, Lee Lindblom, Carlos Lousto, Hiroyuki Nakano, Harald
Pfeiffer, Luciano Rezzolla, and Erik Schnetter.

\appendix*

\section{Post-Newtonian waveforms in the adiabatic approximation}
\renewcommand{\theequation}{A.\arabic{equation}}

There are two basic elements to obtaining a post-Newtonian waveform:
(1) finding the orbital phase of the binary, and (2) using that phase
to find the so-called ``amplitude'' of the waveform.  Previous
references have been incomplete, or predate recent errata involving
spin terms \cite{Blanchet:2006gy,Arun:2009}.  Here, we gather together the most complete and current
formulas for phase and amplitude when spins are \emph{non-precessing}
 (i.e. spins aligned or anti-aligned with the
  orbital angular momentum), using consistent
notation.

\subsection{Phasing}
The orbital phase evolution $\Phi(t)$ of the binary can be computed by
considering energy conservation.  The energy of the full system is
accounted for in three parts, each computed as a function of the
post-Newtonian expansion parameter
\begin{equation}
  \label{eq:DefinevParameter}
  v \define \left( M \frac{d\Phi}{dt} \right)^{1/3}~.
\end{equation}
The first part is the kinetic and gravitational binding energy of the
binary---the orbital energy $E(v)$.  As the system evolves, it gives
off energy to infinity in the form of gravitational waves accounted
for as the flux $\Flux(v)$ leaving the system.  Finally, we must also
account for the tide raised on each black hole by the other, and the
flow of energy into the black holes due to the motion of these tides,
given as a rate of change in the mass of the black holes $\dot{M}(v)$.
Using this threefold accounting for the energy, we can express the
conservation of energy as
\begin{equation}
  \label{eq:EnergyConservation}
  \frac{dE}{dt} + \Flux + \dot{M} = 0~.
\end{equation}
Now, because the expression for the orbital energy is written in terms
of $v$, we can straightforwardly differentiate to find $E'(v)$.  With
the chain rule, $dE/dt = E'(v)\, dv/dt$, we can rearrange this into a
differential equation for $v$:
\begin{equation}
  \label{eq:BalanceEquation}
  \frac{dv}{dt} = - \frac{\Flux(v) + \dot{M}(v)} {E'(v)}~.
\end{equation}
Given the expressions for $\Flux(v)$, $\dot{M}(v)$, and $E'(v)$, this
equation can be integrated to find $v(t)$.  Then, using the definition
of $v$, we see that
\begin{equation}
  \label{eq:PhaseFormula}
  \frac{d\Phi}{dt} = \frac{v^{3}}{M}~,
\end{equation}
which can be integrated in turn to find $\Phi(t)$.  We now exhibit the
formulas for $\Flux(v)$, $\dot{M}(v)$, and $E'(v)$, and discuss
various methods for integrating the balance
equation~\eqref{eq:BalanceEquation}.

Given the masses $M_1$ and $M_2$ and spin vectors $\mathbf{S}_1$ and
$\mathbf{S}_2$, we define the following parameters:
\begin{align}
  \label{eq:SSigma}
  M &\define M_1 + M_2~, \\
  \eta  &\define M_1 \, M_2 / M^2~, \\
  \delta  &\define (M_1 - M_2) / M~,\\
  \bm{\chi}_i &\define \mathbf{S}_i/M_i^2~, \\
  \bm{\chi}_s &\define (\bm{\chi}_1 + \bm{\chi}_2)/2~, \\
  \bm{\chi}_a &\define (\bm{\chi}_1 - \bm{\chi}_2)/2~.
\end{align}
We also define the quantities $\chi_s$ and $\chi_a$ to be the
components of the spin vectors perpendicular to the orbital plane,
namely $\chi_s\define \bm{\chi}_s \cdot \bm{\ell}$ and $\chi_a \define
\bm{\chi}_a \cdot \bm{\ell}$, where $\bm{\ell}$ is the unit vector
along the Newtonian angular momentum.

The orbital energy function can be written in terms of the PN
expansion parameter $v$ defined above as~\cite{Blanchet:2002,
  Blanchet:2005a, Blanchet:2005b, Arun:2009, Blanchet:2006gy,
  Blanchet:2007, Blanchet:2010}\footnote{The 1.5PN and 2.5PN spin
  terms were taken from Eq.~(7.9) of \cite{Blanchet:2006gy}. The 2PN
  spin term and all nonspinning terms were taken from Eq.~(C4) of
  \cite{Arun:2009}.  Note that Eq.~(C5) in the original published
  version of \cite{Arun:2009} is erroneous.}
\begin{equation}\label{eq:Energy}
  \begin{split}
    E(v) & = -\frac{M \eta v^2}{2}\, \left\{1 + v^2\left( -
\frac{3}{4}
        - \frac{\eta}{12} \right) + v^{3} \left[\frac{8 \, \delta
          \chi_a}{3}+\left(\frac{8}{3}
          -\frac{4 \eta }{3}\right) \chi_s \right] \right. \\
    &\qquad\qquad\quad + v^{4} \left[-2 \delta \chi_a
      \chi_s-\frac{\eta ^2}{24}
      +(4 \eta -1) \chi_a^2+\frac{19 \eta }{8}-\chi_s^2-\frac{27}{8}\right] \\
    &\qquad\qquad\quad + v^{5} \left[\chi_a \left(8 \, \delta
        -\frac{31 \delta \eta }{9}\right)
      +\left(\frac{2 \eta^2}{9}-\frac{121 \eta }{9}+8\right) \chi_s \right] \\
    &\qquad\qquad\quad \left. + v^{6} \left[-\frac{35 \eta
          ^3}{5184}-\frac{155 \eta ^2}{96}
        +\left(\frac{34445}{576}-\frac{205 \pi^2}{96}\right) \eta
        -\frac{675}{64} \right] \right\}.
  \end{split}
\end{equation}
We simply take the derivative of this formula with respect to $v$ to
find the energy function appearing in the phasing formula:
\begin{equation}\label{eq:dEnergy}
  \begin{split}
    E'(v) & = -M \eta v\, \left\{ 1 + v^2\left( - \frac{3}{2} -
        \frac{\eta}{6} \right) + v^{3} \left[ \frac{20 \delta
          \chi_a}{3}+\left(\frac{20}{3}
          -\frac{10 \eta }{3}\right) \chi_s \right] \right. \\
    & \hphantom{\quad -\eta v\, \Bigg\{} + v^{4} \left[ -6 \, \delta
      \chi_a \chi_s-\frac{\eta ^2}{8}
      +(12 \eta -3) \chi_a^2+\frac{57 \eta }{8}-3 \chi_s^2-\frac{81}{8} \right]\\
    & \hphantom{\quad -\eta v\, \Bigg\{} + v^{5} \left[ \chi_a
      \left(28 \delta -\frac{217 \delta \eta }{18}\right)
      +\left(\frac{7 \eta^2}{9}-\frac{847 \eta }{18}+28\right) \chi_s \right] \\
    & \hphantom{\quad -\eta v\, \Bigg\{} \left. + v^{6} \left[
        -\frac{35 \eta ^3}{1296}-\frac{155 \eta ^2}{24}
        +\left(\frac{34445}{144}-\frac{205 \pi^2}{24}\right) \eta
        -\frac{675}{16} \right] \right\}~.
  \end{split}
\end{equation}
Similarly, the flux function can be written as~\cite{Blanchet:2002,
  Blanchet:2005a, Blanchet:2005b, Arun:2009, Blanchet:2006gy,
  Blanchet:2007, Blanchet:2010}\footnote{The 1.5PN and 2.5PN spin
  terms were taken from Eq.~(7.11) of \cite{Blanchet:2006gy}. The 2PN
  spin term and all nonspinning terms were taken from Eq.~(C10) of
  \cite{Arun:2009}, except that the term $\eta
  \left\{-\frac{103}{48}(\bm{\chi}_s^2 - \bm{\chi}_a^2) +
    \frac{289}{48} [(\bm{\chi}_{s} \cdot \bm{\ell})^{2} -
    (\bm{\chi}_{a} \cdot \bm{\ell})^{2}] \right\}$ is omitted.  (The
  authors of \cite{Arun:2009} have confirmed that this term should not
  be present.)  Also note that Eq.~(C11) in the original published
  version of \cite{Arun:2009} is erroneous.}
\begin{equation}\label{fluxx}
  \begin{split}
    \Flux(v) & = \frac{32}{5}\, v^{10}\, \eta^2 \left\{ 1 + v^2 \left(
        - \frac{1247}{336} - \frac{35}{12} \eta \right) + v^3
      \left[-\frac{11 \delta \chi_a}{4}+\left(3 \eta
          -\frac{11}{4}\right)
        \chi_s+4 \pi \right] \right.\\
    &\qquad\qquad\qquad + v^4 \left[\frac{33 \delta \chi_a
        \chi_s}{8}+\frac{65 \eta^2}{18} +\left(\frac{33}{16}-8 \eta
      \right) \chi_a^2+\left(\frac{33}{16} -\frac{\eta }{4}\right)
      \chi_s^2
      +\frac{9271 \eta}{504}-\frac{44711}{9072} \right] \\
    &\qquad\qquad\qquad + v^5 \left[\left(\frac{701 \delta \eta }{36}
        -\frac{59 \delta }{16}\right) \chi_a +\left(-\frac{157 \eta
          ^2}{9} +\frac{227 \eta }{9}-\frac{59}{16}\right) \chi_s
      -\frac{583 \pi  \eta }{24}-\frac{8191 \pi }{672} \right] \\
    &\qquad\qquad\qquad + v^6 \left[-\frac{1712}{105} \ln (4
      v)-\frac{1712 \gamma }{105} -\frac{775 \eta ^3}{324}-\frac{94403
        \eta ^2}{3024}+\left(\frac{41 \pi
          ^2}{48}-\frac{134543}{7776}\right) \eta +\frac{16 \pi^2}{3}
      +\frac{6643739519}{69854400} \right] \\
    &\qquad\qquad\qquad \left.  + v^7 \left[\frac{193385 \pi \eta
          ^2}{3024} +\frac{214745 \pi \eta }{1728}-\frac{16285 \pi
        }{504} \right] \right\}~,
  \end{split}
\end{equation}
where $\gamma$ is the Euler Gamma.

Alvi~\cite{Alvi:2001} derived an expression for the transfer of energy
from the orbit to each black hole by means of tidal heating.  The
calculation involves computing the deformation of each hole's horizon
due to the Newtonian field of the other, then using that expression in
formulas for energy absorption due to tidal deformation.  In
particular, his expression is applicable in the comparable-mass case.
By combining the rates of mass change for both black holes, we obtain
the total rate of change:
\begin{equation}
  \label{eq:MassRateOfChange}
  \begin{split}
    \dot{M}(v) &= \frac{32}{5} v^{10} \eta^{2} \left\{
      -\frac{v^{5}}{4}\, \Big[ (1 - 3\eta) \chi_{s} (1 + 3\chi_{s}^{2}
      + 9\chi_{a}^{2}) + (1 - \eta) \delta \chi_{a} (1 + 3\chi_{a}^{2}
      + 9\chi_{s}^{2}) \Big] \right\}~.
  \end{split}
\end{equation}
The coefficient above is the leading-order term in the flux, meaning
that this term is comparable to a relative 2.5PN spin effect in the
flux.  A similar calculation has been carried out in the
extreme-mass-ratio limit~\cite{Tagoshi:1997}, and agrees with this
formula in that limit.  Note that higher-order spin terms were
calculated in~\cite{Alvi:2001}, but are not included here, as they are
at relative 3.5PN order, which is higher than the relative 2.5PN order
to which other spin terms are known.  Except for its explicit presence
in the balance equation~\eqref{eq:BalanceEquation}, we always treat
the mass as a constant.  This leads to additional errors at the 3.5PN
spin level, which we ignore.

Below we will define some of the standard variants of computing
the post-Newtonian phase from the energy and flux functions,
using the naming convention of \cite{Damour:2000zb}.

\subsubsection{TaylorT1 phasing}
The TaylorT1 approximant is computed by numerically integrating the
ordinary differential equation for $v(t)$ in
Eq.~\eqref{eq:BalanceEquation}, using the expressions for orbital
energy, flux, and mass change given in Eqs.~\eqref{eq:dEnergy},
\eqref{fluxx}, and~\eqref{eq:MassRateOfChange}.  The phase is then
computed using this result for $v(t)$ in Eq.~\eqref{eq:PhaseFormula}.

\subsubsection{TaylorT4 phasing}
The TaylorT4 approximant is similar to the TaylorT1 approximant,
except that the ratio of the polynomials on the right-hand side of
Eq.~\eqref{eq:BalanceEquation} is first expanded as a Taylor series,
and truncated at consistent PN order.  Explicitly, the formula to be
integrated is
\begin{equation}\label{eq:dvByDt}
  \begin{split}
    \frac{dv}{dt} & = \frac{32}{5M} v^9 \eta \left\{1 + v^2
      \left[-\frac{11 \eta }{4}-\frac{743}{336} \right] + v^3
      \left[-\frac{113 \delta \chi_a}{12}+\left(\frac{19 \eta }{3}
          -\frac{113}{12}\right) \chi_s+4 \pi \right] \right. \\
    & \qquad\qquad\quad + v^4 \left[\frac{81 \delta \chi_a
        \chi_s}{8}+\frac{59 \eta ^2}{18} +\left(\frac{81}{16}-20 \eta
      \right) \chi_a^2+\left(\frac{81}{16}-\frac{\eta }{4}\right)
      \chi_s^2+\frac{13661 \eta
      }{2016}+\frac{34103}{18144} \right] \\
    & \qquad\qquad\quad + v^5 \left[-\frac{189 \pi \eta
      }{8}-\frac{4159 \pi }{672} + \left(\frac{3 \eta
        }{4}-\frac{3}{4}\right) \delta \chi_a^3 + \left(\frac{9 \eta
        }{4}-\frac{9}{4}\right) \delta \chi_a \chi_s^2
      + \left(\frac{1165 \eta }{24}-\frac{31571}{1008}\right) \delta \chi_a \right. \\
    & \qquad\qquad\qquad \qquad \left. + \left(-\frac{79 \eta ^2}{3}+
        \frac{27 \eta \chi_a^2}{4} -\frac{9 \chi_a^2}{4}+\frac{5791
          \eta }{63}-\frac{31571}{1008}\right) \chi_s
      +\left(\frac{9 \eta }{4}-\frac{3}{4}\right) \chi_s^3 \right] \\
    & \qquad\qquad\quad + v^6 \left[-\frac{1712 \gamma }{105}
      -\frac{5605 \eta ^3}{2592}+\frac{541 \eta
        ^2}{896}+\left(\frac{451 \pi ^2}{48}
        -\frac{56198689}{217728}\right) \eta +\frac{16 \pi ^2}{3}
      +\frac{16447322263}{139708800} \right. \\
    & \qquad\qquad\qquad \qquad -\frac{1712 \ln (4v)}{105} +
    \left(\frac{1517 \eta ^2}{72}-\frac{23441 \eta }{288}
      +\frac{128495}{2016} \right) \chi_s^2 + \left(\frac{565 \delta
        ^2}{9}+\frac{89 \eta ^2}{3}-\frac{2435 \eta }{224}
      +\frac{215}{224} \right) \chi_a^2 \\
    & \qquad\qquad\qquad \qquad \left. + \left(\left(\frac{128495
            \delta }{1008} -\frac{12733 \delta \eta }{144}\right)
        \chi_a +\frac{40 \pi \eta }{3}
        -\frac{80 \pi }{3}\right) \chi_s -\frac{80 \pi  \delta \chi_a}{3} \right] \\
    & \qquad\qquad\quad + v^7 \left[\frac{91495 \pi \eta ^2}{1512}
      +\frac{358675 \pi  \eta }{6048}-\frac{4415 \pi }{4032} \right. \\
    & \qquad\qquad\qquad \qquad \left.  + \left(-\frac{11 \eta ^2}{24}
        +\frac{979 \eta }{24}-\frac{505}{8} \right) \chi_s^3 +
      \left(\frac{\delta \eta ^2}{8}+\frac{742 \delta \eta }{3}
        -\frac{505 \delta }{8} \right) \chi_a^3 \right.\\
    & \qquad\qquad\qquad \qquad \left. + \left(\left(\frac{3 \eta
            ^2}{8}+\frac{917 \eta }{12}
          -\frac{1515}{8}\right) \delta \chi_a + 12 \pi \right) \chi_s^2 \right. \\
    & \qquad\qquad\qquad \qquad \left.  + \left( \left(-124 \delta ^2
          -\frac{3397 \eta ^2}{24}+\frac{7007 \eta}{24}
          -\frac{523}{8}\right) \chi_s -48 \pi  \eta +12 \pi \right) \chi_a^2 \right. \\
    & \qquad\qquad\qquad \qquad \left. + \left(\frac{2045 \eta
          ^3}{216} -\frac{398017 \eta ^2}{2016} +\frac{10772921
          \eta}{54432} -\frac{2529407}{27216}\right) \chi_s
      + 24 \pi  \delta  \chi_a \chi_s \right. \\
    & \qquad\qquad\qquad \qquad \left. \left. + \left(-\frac{41551
            \delta \eta ^2}{864} +\frac{845827 \delta
            \eta}{6048}-\frac{2529407 \delta }{27216}\right) \chi_a
      \right] \right\}~.
  \end{split}
\end{equation}
Note that this expression does not include \emph{all} of the
spin-dependent terms at 3PN and 3.5PN, since the spin terms in the
energy and flux functions are known only up to 2PN and 2.5PN,
respectively.  However, the 3PN and 3.5PN terms shown here will still
be present in this formula when the higher-order terms are included in
the energy and flux formulas.

\subsubsection{TaylorT2 phasing}

Expanding the inverse of Eq.~\eqref{eq:BalanceEquation} allows for
the analytical integration of $t(v)$. The result reads
\begin{align} \label{eq:T2time}
\begin{split}
 t(v) &=  t_0 -  \frac{5M}{256 \eta \, v^8} \Bigg \{ 1 + v^2 \left[
\frac{11 \eta }{3}+\frac{743}{252} \right] + 
v^3 \left[ -\frac{32 \pi}{5} + \frac{226 \delta  \chi_a}{15} + \left(
 \frac{226}{15} -\frac{152 \eta}{15} \right) \chi_s
\right] \\
& \quad + v^4 \left[ 
\frac{3058673}{508032} + \frac{5429 \eta }{504} + \frac{617 \eta
^2}{72} -\frac{81}{4} \delta  \chi _a \chi _s
- \left(\frac{81}{8} - \frac{\eta}{2} \right) \chi_s^2 -
\left(\frac{81}{8} - 40 \eta \right)\chi _a^2 \right] \\
& \quad + v^5 \left[
-\frac{7729\pi }{252} - \frac{13 \pi  \eta }{3} + 
\left( \frac{147101}{756} -\frac{4906 \eta }{27}-\frac{68 \eta
^2}{3}\right) \chi_s
+ \left( \frac{147101
}{756} + \frac{26 \eta }{3} \right) \delta \chi_a \right. \\
& \qquad +  (6- 6\eta) \delta \chi_s^2 \chi_a + (6 - 18 \eta)
\chi_s \chi_a^2 + (2- 6 \eta) \chi_s^3 + (2 - 2\eta) \delta \chi_a^3 
\Bigg] \\
& \quad + v^6 \left[  \frac{6848\gamma}{105} -
\frac{10052469856691}{23471078400} +\frac{128 \pi ^2}{3}
+\left(\frac{3147553127 }{3048192} -\frac{451 \pi ^2}{12}
\right) \eta -\frac{15211 \eta ^2}{1728} + \frac{25565 \eta ^3}{1296} 
\right. \\
& \qquad  + \frac{6848 \ln (4v)}{105} - \left( \frac{584 \pi }{3} -
\frac{448 \pi  \eta }{3} \right) \chi_s -\frac{584 \pi  \delta \,
\chi_a }{3} + \left( \frac{6845}{672} -\frac{43427 \eta }{168} +
\frac{245 \eta ^2}{3} \right) \chi_s^2 \\
& \qquad + \left. \left(  \frac{6845}{672} -\frac{1541 \eta }{12} +
\frac{964 \eta ^2}{3} \right) \chi_a^2 
+ \left( \frac{6845}{336}-\frac{2077 \eta }{6} \right) \delta \,
\chi_s \chi_a
\right] \\
& \quad + v^7 \left[  -\frac{15419335 \pi }{127008} -\frac{75703 \pi 
\eta }{756} + \frac{14809 \pi  \eta ^2}{378} 
+ \left(\frac{4074790483}{1524096} +\frac{30187 \eta }{112}
-\frac{115739 \eta ^2}{216} \right) \delta \, \chi_a
\right. \\
& \qquad + \left( 
\frac{4074790483}{1524096} -\frac{869712071 \eta }{381024}
-\frac{2237903 \eta ^2}{1512} + \frac{14341 \eta ^3}{54}
\right) \chi_s + \left( 228 \pi -16 \pi  \eta \right) \chi_s^2 \\
& \qquad + \left( 228 \pi -896 \pi  \eta \right) \chi_a^2 + 456 \pi 
\delta \, \chi_s \chi_a - \left( \frac{3237}{14} - \frac{14929 \eta
}{84} +  \frac{362 \eta ^2}{3}\right) \chi_s^3 \\
& \qquad - \left( \frac{3237}{14} - \frac{87455 \eta }{84} + 34 \eta
^2 \right) \delta \chi_a^3 - \left( \frac{9711}{14} - \frac{39625 \eta
}{84} +  102 \eta ^2\right) \delta \, \chi_s^2 \chi_a \\
& \qquad \left . - \left(
\frac{9711}{14} -\frac{267527 \eta }{84} + \frac{3574 \eta
^2}{3} \right) \chi_s \chi_a^2 \right] \Bigg\}~.
\end{split}
\end{align}
The comment made below Eq.~\eqref{eq:dvByDt} about spin
contributions at 3PN and 3.5PN order is valid for Eq.~\eqref{eq:T2time}
and the following expansions as well.

The orbital phase $\Phi$ can be integrated similarly to the time $t$. 
Eq.~\eqref{eq:PhaseFormula} and Eq.~\eqref{eq:BalanceEquation} yield
\begin{equation}
 \frac{d\Phi}{dv} = \frac{v^{3}}{M} \, \frac{dt}{dv} = -
\frac{v^{3}}{M} \, \frac{E'(v)}{\Flux(v) + \dot{M}(v)} ~,
\end{equation}
which, after re-expanding in a Taylor series, can be integrated
analytically. The final result reads
{ \allowdisplaybreaks
\begin{align} \label{eq:T2phase}
 \Phi (v) &= \Phi_0 - \frac{1}{32 \eta \, v^5} \Bigg\{
1+ v^2 \left[ \frac{3715}{1008} + \frac{55 \eta }{12} \right] 
+ v^3 \left[ -10 \pi  + \frac{565 \delta  \chi _a}{24} + \left(
 \frac{565}{24} -\frac{95 \eta }{6} \right) \chi_s
\right] \nonumber \\
& \quad + v^4 \left[ \frac{15293365}{1016064} +\frac{27145 \eta
}{1008} + \frac{3085 \eta ^2}{144} -\frac{405 }{8} \delta \,
\chi_a \chi_s - \left( \frac{405}{16} - \frac{5 \eta }{4} \right)
\chi_s^2 - \left( \frac{405}{16} - 100 \eta \right) \chi_a^2 \right]
\nonumber \\
& \quad + v^5 \ln v \left[  \frac{38645 \pi }{672}-\frac{65 \pi  \eta
}{8} - \left( \frac{735505}{2016} - \frac{12265 \eta }{36} - \frac{85
\eta ^2}{2} \right) \chi_s - \left( \frac{735505}{2016}+\frac{65 \eta
}{4} \right) \delta \chi_a  \right. \nonumber \\
& \qquad \left. - \left( \frac{45}{4}-\frac{45 \eta }{4} \right)
\delta \chi_s^2 \chi_a - \left( \frac{45}{4}-\frac{135 \eta }{4}
\right) \chi_s \chi_a^2 - \left( \frac{15}{4}-\frac{45 \eta }{4}
\right) \chi_s^3 - \left( \frac{15}{4}-\frac{15 \eta }{4} \right)
\delta \chi_a^3 \right]  \\
& \quad + v^6 \left[ \frac{12348611926451}{18776862720}-\frac{1712
\gamma}{21}-\frac{160 \pi ^2}{3}-\left( \frac{15737765635 
}{12192768} - \frac{2255 \pi ^2}{48} \right) \eta +\frac{76055 \eta
^2}{6912}-\frac{127825 \eta ^3}{5184} \right. \nonumber \\
& \qquad -\frac{1712 \ln{(4v)}}{21} + \left(\frac{730 \pi
}{3}-\frac{560 \pi  \eta }{3} \right) \chi_s + \frac{730 \pi \delta
\chi_a }{3} -
\left( \frac{34225}{2688}-\frac{217135 \eta }{672}+\frac{1225 \eta
^2}{12} \right) \chi_s^2  \nonumber \\
& \qquad - \left. \left( \frac{34225}{2688}-\frac{7705 \eta
}{48}+\frac{1205 \eta ^2}{3} \right) \chi_a^2 - \left(
\frac{34225}{1344}-\frac{10385 \eta }{24} \right) \delta \, \chi_s
\chi_a \right] \nonumber \\
& \quad + v^7 \left[ \frac{77096675 \pi }{2032128}+\frac{378515 \pi 
\eta }{12096}-\frac{74045 \pi  \eta ^2}{6048} - \left( 
\frac{20373952415}{24385536}+\frac{150935 \eta }{1792}-\frac{578695
\eta ^2}{3456}\right) \delta \chi_a
\right. \nonumber \\
& \qquad - \left( \frac{20373952415}{24385536}-\frac{4348560355 \eta
}{6096384}-\frac{11189515 \eta ^2}{24192}+\frac{71705 \eta ^3}{864}
\right) \chi_s - \left( \frac{285 \pi }{4}-5 \pi  \eta \right)
\chi_s^2 \nonumber \\
& \qquad - \left( \frac{285 \pi }{4}-280 \pi  \eta \right) \chi_a^2 -
\frac{285 \pi }{2} \delta \, \chi_s \chi_a + \left(
\frac{16185}{224}-\frac{74645 \eta }{1344}+\frac{905 \eta ^2}{24}
\right) \chi_s^3 \nonumber \\
& \qquad + \left( \frac{16185}{224}-\frac{437275 \eta }{1344}+\frac{85
\eta ^2}{8} \right) \delta \chi_a^3 + \left(
\frac{48555}{224}-\frac{198125 \eta }{1344}+\frac{255 \eta ^2}{8}
\right) \delta \, \chi_s^2 \chi_a \nonumber \\
& \qquad + \left. \left(  \frac{48555}{224}-\frac{1337635 \eta
}{1344}+\frac{8935 \eta ^2}{24} \right) \chi_s \chi_a^2 \right]
\Bigg\}. \nonumber
\end{align}}
Eq.~\eqref{eq:T2time} and Eq.~\eqref{eq:T2phase} together define
$\Phi(t)$ implicitly.

\subsubsection{TaylorF2 phasing}

Starting from the explicit expressions for time and orbital phase in
the TaylorT2 approximant, it is possible to analytically construct the
Fourier transform of the GW strain in the
framework of the stationary phase approximation (SPA)
\cite{Damour:2000zb,Damour:2002kr,Arun:2004hn}. Denoting the Fourier
transform of Eq.~(\ref{eq:strain}) by $\tilde A_{\ell m} e^{i
\psi_{\ell m}}$, the phase in the frequency domain can be approximated
by
\begin{equation}
 \psi_{\ell m}(f) = 2 \pi f \, t_f - m \Phi(t_f) - \frac{\pi}{4}. 
\end{equation}
Here, $f$ is the Fourier variable and $t_f$ corresponds to the time
when the instantaneous GW frequency coincides with $f$, i.e.,
\begin{equation}
 \frac{d (m \Phi)}{dt} (t_f) = 2 \pi f \qquad \Rightarrow \quad
v(t_f) = \left( \frac{2 \pi M f}{m} \right)^{1/3}. \label{eq:vfreq}
\end{equation}
The form of the Taylor series of $\psi_{\ell m}$ obviously depends on
the spherical harmonic mode's $m$. For the sake of brevity, only the
expansion
for $m=2$ is given below.
{\allowdisplaybreaks
\begin{align}
 \psi_{\ell 2} (v) &= 2 t_0 v^3 - 2\Phi_0 - \frac{\pi}{4} +
\frac{3}{128 \eta \, v^5} \Bigg\{ 1 + v^2
\left[\frac{3715}{756}+\frac{55 \eta }{9}  \right] + v^3 \left[ 
 -16 \pi +\left( \frac{113}{3}-\frac{76 \eta }{3} \right) \chi_s
+\frac{113 \delta  \chi _a}{3} \nonumber \right] \\
& \quad + v^4 \left[ \frac{15293365}{508032}+\frac{27145 \eta
}{504}+\frac{3085 \eta ^2}{72} - \left( \frac{405}{8}-\frac{5 \eta
}{2} \right) \chi_s^2 - \left( \frac{405}{8}-200 \eta \right) \chi_a^2
-\frac{405  }{4} \delta \, \chi_s \chi_a \right] \nonumber \\
& \quad + v^5 \left( 1+ 3  \ln v \right) \left[ \frac{38645
\pi }{756}-\frac{65 \pi  \eta }{9} - \left(
\frac{735505}{2268}-\frac{24530 \eta }{81}-\frac{340 \eta ^2}{9}
\right) \chi_s - \left( \frac{735505}{2268}+\frac{130 \eta }{9}
\right) \delta \chi_a \nonumber \right. \\
& \qquad - \left. (10-10 \eta) \delta \, \chi_s^2 \chi_a - (10-30
\eta) \chi_s \chi_a^2 - \left( \frac{10}{3}-10 \eta \right) \chi_s^3 -
\left( \frac{10}{3}-\frac{10 \eta }{3} \right) \delta \chi_a^3
\right] \nonumber \\
& \quad + v^6 \left[  \frac{11583231236531}{4694215680}-\frac{6848
\gamma}{21}-\frac{640 \pi ^2}{3}-\left(
\frac{15737765635}{3048192}-\frac{2255 \pi ^2}{12} \right) \eta
+ \frac{76055 \eta ^2}{1728} - \frac{127825 \eta ^3}{1296} 
\right. \nonumber \\
& \qquad -\frac{6848  \ln(4v)}{21} + \left( \frac{2920 \pi
}{3}-\frac{2240 \pi  \eta }{3} \right) \chi_s + \frac{2920 \pi 
}{3}\delta \chi_a - \left( \frac{34225}{672}-\frac{217135 \eta
}{168}+\frac{1225 \eta ^2}{3} \right) \chi_s^2 \nonumber \\
& \qquad - \left. \left( \frac{34225}{672}-\frac{7705 \eta
}{12}+\frac{4820 \eta ^2}{3} \right) \chi_a^2 
- \left( \frac{34225}{336}-\frac{10385 \eta }{6} \right) \delta \,
\chi_s \chi_a
\right] \nonumber \\
& \quad + v^7 \left[  \frac{77096675 \pi }{254016}+\frac{378515 \pi 
\eta }{1512}-\frac{74045 \pi  \eta ^2}{756} -
\left( \frac{20373952415}{3048192}+\frac{150935 \eta
}{224}-\frac{578695 \eta ^2}{432} \right) \delta \chi_a
\right. \nonumber \\
& \qquad - \left( \frac{20373952415}{3048192}-\frac{4348560355 \eta
}{762048}-\frac{11189515 \eta ^2}{3024}+\frac{71705 \eta ^3}{108} 
\right) \chi_s - (570 \pi -40 \pi  \eta) \chi_s^2 \nonumber \\
& \qquad - (570 \pi -2240 \pi  \eta) \chi_a^2 -1140 \pi  \delta \,
\chi_s \chi_a + \left( \frac{16185}{28}-\frac{74645 \eta
}{168}+\frac{905 \eta ^2}{3} \right) \chi_s^3 \nonumber \\
& \qquad + \left( \frac{16185}{28}-\frac{437275 \eta }{168}+85 \eta ^2
\right) \delta \chi_a^3 + \left( \frac{48555}{28}-\frac{198125 \eta
}{168}+255 \eta ^2 \right) \delta \, \chi_s^2 \chi_a \nonumber \\
& \qquad \left. + \left( \frac{48555}{28}-\frac{1337635 \eta
}{168}+\frac{8935 \eta ^2}{3} \right) \chi_s \chi_a^2  \right] \Bigg
\}.
\end{align}
}%
According to Eq.~(\ref{eq:vfreq}), $v$ should be understood as $v =
(M \pi f)^{1/3}$ in the equation above.

\subsection{Waveform amplitudes}
\label{sec:WaveformAmplitudes}
Now, given the orbital phase $\Phi$ and the related post-Newtonian
expansion parameter $v$ defined in Eq.~\eqref{eq:DefinevParameter}, we
can obtain the waveform observed at infinity.  Currently, the most complete
expressions for the nonspinning parts of the waveform are
found in~\cite{Blanchet:2008}.  In particular, Eqs.~(9.3) and~(9.4) of
that reference give the decomposition of $h_+ - ih_\times$ into
harmonics as requested in Eq.~\eqref{eq:hDecomposition}.  Due to space
considerations and the danger of transcription errors, we do not
reproduce those equations here, but simply refer the reader to that
paper.  To these, we must add\footnote{Note that
  Refs.~\cite{Blanchet:2008} and~\cite{WillWiseman:1996} share the
  notation set forth in~\cite{Blanchet:1996}, so that we can simply
  add the relevant terms.  Also note that each of those references
  uses a different normalization for the variable $H$ compared to the
  one used here.}  the spin terms given most completely
in~\cite{WillWiseman:1996}.  There, the spin terms were not explicitly
decomposed into harmonics, however, using Eq.~(9.2)
of~\cite{Blanchet:2008}, it is a simple matter to deduce them.  Using
Eqs.~(F24) and (F25) of~\cite{WillWiseman:1996}, and noting the
overall sign error in Eq.~(F25c), we obtain the only nonzero spin
contributions to the harmonics:
\begin{align}
  \label{eq:SpinAmplitudeTerms}
  H_{2,2} &= -\frac{16}{3} \sqrt{\frac{\pi}{5}} v^5 \eta \left[2 \delta
    \chi_{a} + 2(1-\eta) \chi_{s} + 3 v \eta \left(\chi_{a}^2 -
      \chi_{s}^2 \right) \right] e^{-2 i \Phi}~, \\
  H_{2,1} &= 4 i \sqrt{\frac{\pi}{5}} v^4 \eta (\delta \chi_{s} +
  \chi_{a}) e^{-i \Phi}~, \\
  H_{3,2} &= \frac{32}{3} \sqrt{\frac{\pi}{7}} v^5 \eta^2 \chi_{s}
  e^{-2 i \Phi}~.
\end{align}
In all cases, modes with negative values of $m$ can be obtained from
\begin{equation}
  \label{eq:NegativeMModes}
  H_{\ell, -m} = (-1)^{\ell}\, \bar{H}_{\ell,m}~.
\end{equation}

The appropriate SPA amplitude in Fourier space can easily be deduced
from its time-domain description $A_{\ell m}$ by
\begin{equation}
 \tilde A_{\ell m} = A_{\ell m} \sqrt{\frac{2\pi}{m \ddot \Phi}} =
A_{\ell m} \sqrt{\frac{2 \pi M}{3m v^2 \, \dot v}}~,
\end{equation}
where $\dot v$ can be taken for instance from Eq.~(\ref{eq:dvByDt})
and all arguments should be replaced according to
Eq.~(\ref{eq:vfreq}).


\bibliography{NRDataFormat}

\end{document}